\documentclass[prd,superscriptaddress,showpacs,preprintnumbers,amsmath,amssymb]{revtex4}

\usepackage{graphicx}
\usepackage{dcolumn}
\usepackage{bm}
\usepackage{epsfig}
\usepackage{amsfonts}

\begin{document}

\title{Massive particles' Hawking radiation via tunneling from the G.H
Dilaton black hole }

\author{Yapeng Hu}
  \email{huzhengzhong2050@163.com}
  \affiliation{Department of Physics,
Beijing Normal University, Beijing 100875, China}

\author{Jingyi Zhang}
   \email{physicz@263.net}
   \affiliation {Center for Astrophysics,
Guangzhou University, Guangzhou 510006, China}

\author{Zheng Zhao}
   \email{zhaoz43@hotmail.com}
   \affiliation {Department of Physics,
Beijing Normal University, Beijing 100875, China}

\begin{abstract}
In the past, Hawking radiation was viewed as a tunneling process and
the barrier was just created by the outgoing particle itself. In
this paper, Parikh's recent work is extended to the case of massive
particles' tunneling. We investigate the behavior of the tunneling
massive particles from a particular black hole solution-G.H Dilaton
black hole which is obtained from the string theory, and calculate
the emission rate at which massive particles tunnel across the event
horizon. We obtain that the result is also consistent with an
underlying unitary theory. Furthermore, the result takes the same
functional form as that of massless particles.

\end{abstract}
 \maketitle

\section{Introduction}

In 2000, Parikh and Wilczek proposed a new method to calculate the emission
rate at which particles tunnel across the event horizon\cite{1,2,3,4}. The
key points in their method were that they treated Hawking radiation as a
tunneling process and considered that the barrier was just created by the
outgoing particle itself, in particular, they found a coordinate system\
well-behaved at the event horizon to calculate the emission rate\cite{7,8}.
In this way they obtained the corrected emission spectrum of the particles
from the spherically symmetric black holes, such as Schwarzschild black hole
and Reissner-Norstrom black hole. It's found that their results were
consistent with an underlying unitary theory. However, the particles which
they have treated all are massless and follow the radial lightlike geodesics
when they tunnel across the horizon, while the massive particles don't
follow the radial lightlike geodesics. In this paper, we extend Parikh's
method to a particular black hole solution-G.H Dilaton black hole\cite{10},
and investigate the massive particles' tunneling\cite{5,6}. Then, we
calculate the emission rate at the event horizon of the G.H Dilaton black
hole. During the calculation, for the sake of simplicity, we consider the
outgoing massive particle as a massive shell (de Broglie s-wave), and the
phase velocity and group velocity of the de Broglie wave corresponding to
the outgoing particle are respectively obtained.

The rest of the paper is organized as follows. In section 2, we first
introduce a Painleve-G.H Dilaton coordinate system\cite{10}, and then
investigate the behavior of the massive tunneling particles. In section 3,
we calculate the emission rate at which massive particles tunnel across the
event horizon of G.H Dilaton black hole. Finally, in section 4, we give a
brief conclusion and discussion.

\section{Painleve coordinates and the behavior of the massive tunneling
particles}

The G.H Dilaton black hole metric is\cite{9,10}

\begin{equation}
ds^{2}=e^{2U}dt_{s}^{2}-e^{-2U}dr^{2}-R^{2}(r)(d\theta ^{2}+\sin ^{2}\theta
d\varphi ^{2}),  \label{1}
\end{equation}%
where

\begin{eqnarray}
e^{2U} &=&(1-\frac{r_{H}}{r})(1-\frac{r_{-}}{r})^{\frac{1-a^{2}}{1+a^{2}}},
\notag \\
R &=&r(1-\frac{r_{-}}{r})^{\frac{a^{2}}{1+a^{2}}}.  \label{2}
\end{eqnarray}%
and $r_{H}$, $r_{-}$ are respectively the event horizon and the inner
horizon, constant $a$ is a coupling coefficient. The mass and the charge of
the black hole are respectively

\begin{eqnarray}
M &=&\frac{r_{H}}{2}+\frac{1-a^{2}}{1+a^{2}}\frac{r_{-}}{2},  \notag \\
Q^{2} &=&\frac{r_{H}r_{-}}{1+a^{2}}.  \label{3}
\end{eqnarray}

It is easy to find that the metric in (1) is singular at the location of the
event horizon, $r=r_{H}$. As mentioned in sec.1, we should first find a
coordinate system to calculate the emission rate which is well-behaved at
the event horizon. It's found that the Painleve coordinate system is
convenient\cite{11}. If let $dt=dt_{s}+\frac{1}{\Delta }\frac{\sqrt{g}}{1-g}%
dr$, where $g=\frac{r_{H}}{r},\Delta =(1-\frac{r_{-}}{r})^{\frac{1-a^{2}}{%
1+a^{2}}},$ we can read the Painleve-G.H Dilaton line element\cite{10}

\begin{eqnarray}
ds^{2} &=&(1-g)\Delta dt^{2}-2\sqrt{g}dtdr-\frac{1}{\Delta }%
dr^{2}-R^{2}(r)(d\theta ^{2}+\sin ^{2}\theta d\varphi ^{2})  \notag \\
&=&g_{00}dt^{2}+2g_{01}dtdr-\frac{1}{\Delta }dr^{2}-R^{2}(r)(d\theta
^{2}+\sin ^{2}\theta d\varphi ^{2}).  \label{4}
\end{eqnarray}

The metric (4) displays the stationary, nonstatic, and nonsingular nature of
the space time. Moreover, we find another important feature which we will
describe as follows.

As we know, according to Landau's theory of the coordinate clock
synchronization, the difference of coordinate times of two events taking
place simultaneously in different place is\cite{12}%
\begin{equation}
\bigtriangleup T=-\int \frac{g_{0i}}{g_{00}}dx^{i}.  \label{5}
\end{equation}

where i=1,2,3. And we consider the space-time has been decomposed in 3+1.
So, if the simultaneity of coordinate clocks can be transmitted from one
place to another and have nothing to do with the integration path, the
components of the metric should satisfy\cite{15}%
\begin{equation}
\frac{\partial }{\partial x^{j}}(\frac{g_{0i}}{g_{00}})=\frac{\partial }{%
\partial x^{i}}(\frac{g_{0j}}{g_{00}})\text{ \ }(i,j=1,2,3).  \label{6}
\end{equation}

It's easy to find that the line element (4)\ satisfies condition (6). That
is, the coordinate clock synchronization in the Painleve coordinates can be
transmitted from one place to another, though the line element is not
diagonal. In quantum mechanics, it is an instantaneous process when particle
tunnels across a barrier. Thus, this feature is necessary for us to discuss
the tunneling process.

From (4) we can easily obtain the radial null geodesics\cite{10}

\begin{equation}
\overset{\cdot }{r}\equiv \frac{dr}{dt}=-\Delta \sqrt{g}\pm \Delta .
\label{7}
\end{equation}%
where the upper(lower) sign corresponds the outgoing(ingoing) geodesic.

But in this paper we consider the tunneling of massive particles. When the
massive particles tunnel across the horizon, they do not follow the radial
lightlike geodesics in (7). In order to obtain the $\overset{\cdot }{r}$ of
massive particles, and for the sake of simplicity, we consider the outgoing
massive particle as a massive shell (nonrelativistic de Broglie s-wave).
According to the WKB formula, the approximative wave equation is\cite{5,6}

\begin{equation}
\Psi (r,t)=Ce^{i(\int\nolimits_{r_{i}-\varepsilon }^{r}p_{r}dr-\omega t)}.
\label{8}
\end{equation}%
where $r_{i}-\varepsilon $ represents the initial location of the
particle. Using the non-relativistic quantum mechanics, and if we
let

\begin{equation}
\int\nolimits_{r_{i}-\varepsilon }^{r}p_{r}dr-\omega t=\phi _{0},  \label{9}
\end{equation}%
then, we have

\begin{equation}
\frac{dr}{dt}=\overset{\cdot }{r}=\frac{\omega }{k}.  \label{10}
\end{equation}%
where $k$ is the de Broglie wave number. Comparing the definition of the
phase velocity, we find that $\overset{\cdot }{r}$ is just the phase
velocity of the de Broglie wave. For the nonrelativistic de Broglie wave,
the definitions of the group velocity $v_{g}$ and the phase velocity $v_{p}$%
, and the relationship between them are\cite{16}%
\begin{equation}
v_{p}=\frac{dr}{dt}=\overset{\cdot }{r}=\frac{\omega }{k},  \label{11}
\end{equation}%
\begin{equation}
v_{g}=\frac{dr_{c}}{dt}=\frac{d\omega }{dk},  \label{12}
\end{equation}%
\begin{equation}
v_{p}=\frac{1}{2}v_{g}.  \label{13}
\end{equation}%
where $r_{c}$ is the location of the particle. In order to obtain the
formula of the phase velocity $\overset{\cdot }{r}$, let us first
investigate the behavior of a massive particle tunneling across the horizon%
\cite{5,6}.

Since tunneling across the barrier is an instantaneous process, there are
two simultaneous events during the process of emission. One event is
particle tunneling into the barrier, and the other is particle tunneling out
the barrier. In terms of Landau's theory of the coordinate clock
synchronization, the difference of coordinate times of these two
simultaneous events is

\begin{equation}
dt=-\frac{g_{0i}}{g_{00}}dx^{i}=-\frac{g_{01}}{g_{00}}dr_{c}\ \text{\ }%
(d\theta =d\varphi =0),  \label{14}
\end{equation}%
so the group velocity is

\begin{equation}
v_{g}=\frac{dr_{c}}{dt}=-\frac{g_{00}}{g_{01}},  \label{15}
\end{equation}%
and therefore the phase velocity is

\begin{equation}
v_{p}=\overset{\cdot }{r}=\frac{1}{2}v_{g}=-\frac{1}{2}\frac{g_{00}}{g_{01}}.
\label{16}
\end{equation}%
Substituting $g_{00}$ and $g_{01}$ into (16), we obtain the expression of $%
\overset{\cdot }{r}$%
\begin{equation}
\overset{\cdot }{r}=-\frac{1}{2}\frac{g_{00}}{g_{01}}=\frac{1}{2}\frac{%
(1-g)\Delta }{\sqrt{g}}.  \label{17}
\end{equation}%
Here (17) is corresponding to the outgoing motion of the massive particles,
and we don't consider the effect of self-gravitation. If the
self-gravitation is included, (17) should be modified by replacing $M$ with $%
M-\omega $, where $\omega $ is the particle's energy.

\section{Massive particles' tunneling and the emission rate}

For a positive-energy s-wave, the rate of tunneling $\Gamma $ could take the
form

\begin{equation}
\Gamma \sim \exp (-2\text{Im}I).  \label{18}
\end{equation}%
where $I(r)$ is the action. According to the WKB approximation, the action
has been found to have a conveniently simple form\cite{1}

\begin{equation}
I=\int_{r_{i}}^{\ r_{f}}p_{r}dr=\int\nolimits_{r_{i}}^{\
r_{f}}\int_{0}^{p_{r}}dp_{r}^{^{\prime }}dr.  \label{19}
\end{equation}%
where $p_{r}$ is the radial momentum. And $r_{i}$ is the initial radius
corresponding the site of pair-creation, which should be slightly inside the
event horizon $r_{H}$, while $r_{f}$ is the final radius, which is slightly
outside the final position of the horizon.

Using the Hamilton's equation $\frac{dH}{dp_{r}}=\overset{\cdot }{r}$, and
substituting (17) into (19), the action will be

\begin{equation}
I=\int_{r_{i}}^{\ r_{f}}p_{r}dr=\int\nolimits_{r_{i}}^{\
r_{f}}\int_{M_{i}}^{M_{f}}\frac{dM}{\overset{\cdot }{r}}dr=%
\int_{M_{i}}^{M_{f}}\int\nolimits_{r_{i}}^{\ r_{f}}\frac{2\sqrt{rr_{H}}drdM}{%
(r-r_{H})(1-\frac{r_{-}}{r})^{\frac{1-a^{2}}{1+a^{2}}}}.  \label{20}
\end{equation}%
where the Hamiltonian $H=M$, $M_{i}$ and $M_{f}$ are respectively the
initial mass and the final mass of the black hole.

From (18), we get that we can just consider the imaginary part of the
action. In (20) if we do the r integral first, and then we can obtain the
imaginary part of the action

\begin{equation}
\text{Im}I=-\pi \int_{M_{i}}^{M_{f}}\frac{2r_{H}}{(1-\frac{r_{-}}{r_{H}})^{%
\frac{1-a^{2}}{1+a^{2}}}}dM=-\pi \int_{M_{i}}^{M_{f}}\frac{2r_{H}{}^{\frac{2%
}{1+a^{2}}}}{(r_{H}-r_{-})^{\frac{1-a^{2}}{1+a^{2}}}}dM.  \label{21}
\end{equation}

In order to calculate (21) conveniently, we can change the variant $M$ into $%
r_{H}$\cite{10,13}, which we can get from (2) (3)

\begin{equation}
dM=\frac{r_{H}^{2}-(1-a^{2})Q^{2}}{2r_{H}^{2}}dr_{H}.  \label{22}
\end{equation}%
Substituting (22) into (21) yields

\begin{equation}
\text{Im}I=\int_{r_{i}}^{r_{f}}k(r_{H})dr_{H},  \label{23}
\end{equation}%
where

\begin{equation}
k(r_{H})=-\pi r_{H}^{\frac{1-3a^{2}}{1+a^{2}}}\frac{r_{H}^{2}-Q^{2}(1-a^{2})%
}{[r_{H}^{2}-Q^{2}(1+a^{2})]^{\frac{1-a^{2}}{1+a^{2}}}}.  \label{24}
\end{equation}

The integral in (23) is not easy to calculate directly. However, what we
need is not the direct result but the comparison between $\text{Im}I$ and $%
\Delta S$. So first we can expand $k(r_{H})$ at the near field of $r_{i}$ as
a Taylor series\cite{10,13}

\begin{equation}
k(r_{H})=k(r_{i})+k^{\prime }(r_{i})(r_{H}-r_{i})+\frac{k^{(2)}(r_{i})}{2}%
(r_{H}-r_{i})^{2}+\cdot \cdot \cdot .  \label{25}
\end{equation}%
Substituting (25) into (23), we obtain

\begin{equation}
\text{Im}I=k(r_{i})\Delta r_{H}+\frac{1}{2}k^{\prime }(r_{i})(\Delta
r_{H})^{2}+\frac{k^{(2)}(r_{i})}{3!}(\Delta r_{H})^{3}+\cdot \cdot \cdot .
\label{26}
\end{equation}%
where $\Delta r_{H}=r_{f}-r_{i}$.

For the G.H Dilaton black hole, the entropy is\cite{10,14}

\begin{eqnarray}
S &=&\frac{1}{4}A=\pi R^{2}(r_{H})=\pi r_{H}^{2}(1-\frac{r_{-}}{r_{H}})^{%
\frac{2a^{2}}{1+a^{2}}}  \notag \\
&=&\pi \lbrack r_{H}^{2}-Q^{2}(1-a^{2})]^{\frac{2a^{2}}{1+a^{2}}}r_{H}^{%
\frac{2(1-a^{2})}{1+a^{2}}}  \label{27}
\end{eqnarray}

The difference of the entropies of the black hole before and after the
emission is

\begin{equation}
\Delta S=S(r_{f})-S(r_{i})=\frac{dS}{dr_{H}}\Delta r_{H}+\frac{1}{2!}\frac{%
d^{2}S}{dr_{H}^{2}}(\Delta r_{H})^{2}+\frac{1}{3!}\frac{d^{3}S}{dr_{H}^{3}}%
(\Delta r_{H})^{3}+\cdot \cdot \cdot .  \label{28}
\end{equation}

Using (27), we get

\begin{equation}
\frac{dS}{dr_{H}}=2\pi r_{H}^{\frac{1-3a^{2}}{1+a^{2}}}\frac{%
r_{H}^{2}-Q^{2}(1-a^{2})}{[r_{H}^{2}-Q^{2}(1+a^{2})]^{\frac{1-a^{2}}{1+a^{2}}%
}}.  \label{29}
\end{equation}

In principle, if we substitute (29) into (28), we can obtain $\Delta S$, and
then we can compare it with (26). But fortunately, comparing (29) with (24),
we find%
\begin{equation}
\frac{dS}{dr_{H}}=-2k(r_{H})  \label{30}
\end{equation}
\ \ \ Now, we can easily obtain the following equation by comparing (26)
with (28) and using (30)

\begin{equation}
\Delta S=-2\text{Im}I.  \label{31}
\end{equation}%
which expresses that the emission rate is also consistent with an underlying
unitary theory\cite{10}.

\section{Conclusion and discussion}

In this paper, the space-time is a spherically symmetric charged dilaton
black hole. Not as usual is that it is obtained from the string theory\cite%
{9}. However, the result is also consistent with the underlying unitary
theory. Comparing with the Schwarschild-desitter space-time in Ref\cite{5},
We can find that the conclusion at the end of Sec.3 is more general.
Moreover,\ the particles which we discuss are massive, but the results take
the same functional form as that of massless particles\cite{10}, which
expresses that the tunneling effect is an intrinsic property of the black
hole. As a further discussion and viewed from the calculation, we find that
though $\overset{\cdot }{r}$ of massive particles in (17) is different from
that of massless particles in (7), the results are the same after we do the
r integral first in (20). They are all equal to

\begin{equation}
\text{Im}I=-\int\nolimits_{M_{i}}^{M_{f}}f(r_{H})dM,\text{ here }f(r_{H})=%
\frac{2\pi r_{H}{}^{\frac{2}{1+a^{2}}}}{(r_{H}-r_{-})^{\frac{1-a^{2}}{1+a^{2}%
}}}.  \label{32}
\end{equation}

For the G.H Dilaton black hole, the expression of the inverse temperature, $%
\beta $, is\cite{10}

\begin{equation}
\beta \equiv \frac{1}{T}=\frac{4\pi r_{H}{}^{\frac{2}{1+a^{2}}}}{%
(r_{H}-r_{-})^{\frac{1-a^{2}}{1+a^{2}}}}.  \label{33}
\end{equation}%
Comparing (32) with (33), we can easily obtain

\begin{equation}
f(r_{H})=\frac{1}{2}\beta .  \label{34}
\end{equation}%
which implicates more clearly that the tunneling effect is an intrinsic
property of the black holes.

\section{Acknowledgements}

This work is supported by the National Natural Science Foundation of
China under Grant Nos.10475013,10373003 and the National Basic
Research Program of China Grant No.2003CB716300. We are grateful to
the referees for the help in improving the article.

\end{document}